\documentclass[aps,prx,reprint,superscriptaddress,longbibliography,floatfix]{revtex4-2}
\usepackage{amsmath,amssymb,bm,graphicx,booktabs}
\usepackage[normalem]{ulem}\usepackage{xcolor}

\newcommand{\Sztot}{{S^z_{\rm tot}}}
\usepackage[colorlinks=true,allcolors=blue]{hyperref}
\graphicspath{{figures/arxiv/}}
\newcommand{\rr}{\mathbf r}
\newcommand{\SSS}{\mathbf S}
\newcommand{\NN}{\mathrm{NN}}
\newcommand{\var}{\operatorname{Var}}
\newcommand{\moire}{moir\'e}
\newcommand{\Moire}{Moir\'e}
\newcommand{\av}[1]{\left\langle #1\right\rangle}
\begin{document}

\title{From continuum neural quantum states to effective lattice models: Fluctuating Wigner molecules in \moire{} superlattices}

\author{Ahmed Khalifa}
\affiliation{Department of Physics, Carnegie Mellon University, Pittsburgh, PA 15213, USA}
\author{Shubhayu Chatterjee}
\affiliation{Department of Physics, Carnegie Mellon University, Pittsburgh, PA 15213, USA}

\begin{abstract}
Effective descriptions of quantum materials rely on a separation of energy scales that
identifies their low-energy degrees of freedom. When kinetic energy, lattice potential, and interactions compete, this hierarchy breaks down, and both the microscopic state and its effective description are unknown.
Continuum neural quantum states (NQS) can find the ground state directly without projecting onto prescribed bands, orbitals, or spins, but their microscopic wavefunctions are difficult to interpret in terms of effective variables.
Here, we introduce a wavefunction readout that maps the continuum NQS onto site-resolved charge and spin statistics, allowing effective degrees of freedom to emerge from the microscopic solution.
Focusing on moir\'e transition metal dichalcogenides at filling $\nu = 3$, the NQS reveals a low-spin ground state below the ferromagnetic crystal of spin-polarized  Wigner molecules.
Although the charge density retains triangular molecular motifs, our readout reveals significant charge fluctuations through bound nearest-neighbor doublon--hole pairs.
Each molecular charge state has an inert spin-singlet core and rim electrons that carry the spin.
Charge and spin fluctuations are therefore intertwined, with antiferromagnetic spin-correlations between sites, motivating an effective description by a spinful two-orbital Hubbard model.
More generally, our results demonstrate a route from continuum wavefunctions to effective lattice models when competing scales preclude a controlled reduction.
\end{abstract}

\maketitle

\section{Introduction}
\label{sec:intro}

Effective theories make quantum materials intelligible by replacing their microscopic electronic degrees of freedom by a small, tractable set of low-energy variables.
Such a reduction is well controlled when there exists a hierarchy of energy scales for the relevant electronic bands, orbitals and local spin multiplets.
However, when the electronic kinetic energy, the lattice potential and the inter-electronic interaction compete on comparable energy scales, that hierarchy is lost.
This poses a significant challenge: one must solve the microscopic many-body problem without a small parameter and determine which local objects and fluctuations constitute its effective description.
In this regime, an effective model cannot be assumed at the outset; it must emerge from the solution.

Recent advances in continuum neural quantum states (NQS) offer a promising route towards addressing this challenge.
NQS constitute a class of variational states that can represent the full many-electron wavefunction in the continuum, without requiring a projection onto a prescribed set of bands, orbitals or spin.
The NQS is then optimized directly against the microscopic Hamiltonian to obtain the ground state.
Indeed, neural wavefunctions, including self-attention architectures~\cite{vonGlehn2023,qxc3-bkc7}, have been applied to obtain ground states of correlated electronic systems~\cite{qxc3-bkc7,GeierFu2026,LuoDaiFu2024,Li2025CommPhys}.
The price, however, is interpretability.
A microscopic solution with millions of variational parameters is not yet an effective description: its energy and one-body density do not reveal the local charge states, orbital and spin moments, and associated correlations that an effective low-energy description must retain.
Extracting this physics thus
requires observables that ask the network precisely the questions an effective model needs answered.

In this work, we devise a statistical readout of continuum NQS that allows us to identify effective local degrees of freedom and construct low-energy lattice Hamiltonians to describe their fluctuations.
Crucially, the network is given no a priori information beyond the continuum Hamiltonian: the effective lattice and its fluctuating degrees of freedom emerge directly from the variational ground state solution.

We illustrate the power and flexibility of our approach by focusing on correlated electronic states realized in moir\'e superlattices of semiconducting transition metal dichalcogenides (TMDs)~\cite{Regan2020,Tang2020,LiT2021,Ghiotto2021}.
This platform provides a generic and minimal realization of an interacting electronic system in a periodic potential without a clear hierarchy of energy scales~\cite{LiT2021cap,Artificial_Atoms,YannouleasLandman2023,Khalifa2026SciPost}, while also offering the prospect of novel correlated phases in light of the intense experimental activity on twisted TMDs~\cite{Mak2022, Zhai2025, LiBohao2026}.
The smooth moir\'e periodic potential reconstructs the quadratically dispersing TMD valence or conduction bands into multiple moir\'e mini-bands.
At small twist angles and at one carrier per moir\'e site ($\nu = 1$), a semi-isolated flat band was proposed to realize a single-orbital Hubbard model on the triangular lattice, and, at strong coupling, its Heisenberg limit~\cite{Wu2018,AngeliMacDonald2021}.
Experiments have observed the corresponding Mott insulators~\cite{Regan2020,Tang2020} and continuous Mott transitions~\cite{LiT2021,Ghiotto2021}. Higher integer fillings offer the possibility of richer correlated physics: several internal states of each moir\'e artificial atom can participate, enlarging
the local Hilbert space and coupling charge, orbital, and spin degrees of freedom.
The same internal structure that makes this regime interesting also removes the obvious route from the continuum system to an effective lattice model.

Filling $\nu=3$ is a concrete realization of this opportunity.
Correlated insulators persist at higher integer fillings~\cite{LiT2021cap}, where the Coulomb repulsion within each moir\'e potential well can well exceed its intrinsic non-interacting level spacing, and the interaction between neighboring wells can compete with the separation between the lowest bands.
When on-site interactions constitute the dominant energy scale, three electrons localize within each well into a tri-lobed Wigner molecule, the zero-dimensional counterpart of a Wigner crystal~\cite{Artificial_Atoms,YannouleasLandman2023,YannouleasLandman2024PRB,YannouleasLandman2024PRL}.
Scanning tunneling microscopy has imaged such molecular charge patterns in twisted WS$_2$~\cite{Li2024Science} and twisted MoTe$_2$~\cite{Liu2026NSR}.
While the spin-density is yet to be measured, the prevailing mean-field picture of the $\nu = 3$ state is a fully spin-polarized Wigner-molecule crystal: three electrons localize in every well and form a lattice of identical triangular molecules~\cite{Artificial_Atoms,Li2024Science}.

Whether this picture survives electronic fluctuations is not known.
Exact treatments of an isolated well determine its molecular states but exclude charge transfer between molecules~\cite{Artificial_Atoms,YannouleasLandman2024PRL,Khalifa2026SciPost}.
Hartree--Fock theory treats the full lattice, but replaces fluctuating moments by static ones and favors a fully polarized ferromagnet ($S = 3/2$) via exchange~\cite{Artificial_Atoms}.
A recent neural-network calculation found a variational state with spin $S = \tfrac12$ per site and lower energy than the fully polarized state~\cite{Li2025CommPhys}, indicating that spin fluctuations can destabilize the mean-field state.
However, an effective model that accurately describes the low-energy charge and spin fluctuations remains elusive.

We find that allowing for spin fluctuations leads to significant charge fluctuations that destabilize the mean-field fully polarized state.
The variational ground state lies in the sector with zero total spin, in sharp contrast with Hartree--Fock predictions~\cite{Artificial_Atoms} and existing NQS results~\cite{LuoDaiFu2024,Li2025CommPhys}.
Remarkably, the network itself first settles into a crystal-like state resembling the mean-field solution, with nearly frozen charges and static moments, before reaching the ground state via local quantum tunneling events that release charge and spin together, one molecule at a time.
To interpret this state, we develop a readout that coarse-grains the wavefunction onto the moir\'e sites.
Each sampled configuration becomes a set of site charges and site spins, whose distributions and correlations are measured directly from the variational state, while spin-exchange amplitude ratios determine the spin correlations.
While the one-body charge density of the low-spin ground state remains a lattice of identical triangular motifs, the underlying wavefunction exhibits substantial fluctuations between molecular charge states with two, three, and four electrons, predominantly through bound nearest-neighbor doublon--hole (4e--2e) pairs.
Additionally, our readout exposes a simple local organization: two electrons form a robust spin-singlet core, and the remaining rim electrons carry the spin.
The charge of a molecule therefore controls its local moment; removing a charge also depletes the spin and vice-versa.
Antiferromagnetic correlations are present in every charge environment and are concentrated on the doublons (4e molecules).
Collectively, these observations identify the minimal effective degrees of
freedom and motivate a two-orbital Hubbard model of the rim electrons above an inert core, with Hund's coupling and a nearest-neighbor interaction treating charge fluctuations and magnetism
on the same footing.

The $\nu=3$ moir\'e system thus illustrates a general route from a microscopic problem with competing scales to an effective theory via neural quantum states: solve the continuum many-body state without choosing its local degrees of freedom in advance, use a readout to expose local statistics encoded in the wavefunction, and build the effective model from the structures those statistics reveal.

\section{\Moire{} Hamiltonian and neural-network wavefunction}
\label{sec:system}

\begin{figure*}[t]
\includegraphics{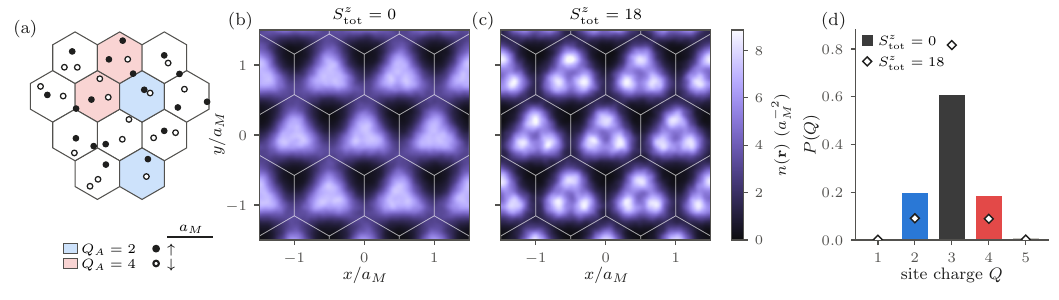}
\caption{Coarse-graining the neural-network wavefunction onto \moire{} sites (reference
point). (a)~A configuration of the 36 electrons sampled from $|\Psi|^2$ at $\Sztot=0$, shown in the
twelve Wigner--Seitz cells of the periodic cell; filled (open) circles denote spin-up
(spin-down) electrons. Assigning each electron to its cell defines the site charges $Q_A$:
shaded cells hold two (blue) or four (red) electrons, while the others hold three. The
configuration is drawn at random from those with the typical number of doublons and holes, with a doublon next to each hole. (b),(c)~The one-body density at $\Sztot=0$ and in the
fully polarized state, $\Sztot=18$, on a common color scale; lines mark the cells. Both consist of
twelve identical triangles; at $\Sztot=18$ the three electrons of each molecule are resolved at
its corners. (d)~The distribution of site charges at $\Sztot=0$ (bars) and $\Sztot=18$ (diamonds).}
\label{fig:cg}
\end{figure*}

We begin by introducing the continuum model and the variational wavefunction. We consider
electrons in a two-dimensional triangular \moire{} potential at filling $\nu=3$. Measuring
lengths in units of the \moire{} lattice constant $a_M$ and energies in units of
$E_0=\hbar^2/(m^\ast a_M^2)$, the Hamiltonian reads
\begin{equation}
\begin{aligned}
H&=\sum_i\left[-\frac12\nabla_i^2+ V(\rr_i)\right]
+\sum_{i<j}\frac{U}{|\rr_i-\rr_j|},\\
 V(\rr)&=-2 V_0\sum_{n=0}^{2}\cos(\mathbf g_n\cdot\rr+\phi),
\end{aligned}
\label{eq:hamiltonian}
\end{equation}
with $\mathbf g_n=(4\pi/\sqrt3)(\sin\tfrac{2\pi n}{3},\cos\tfrac{2\pi n}{3})$. The physics is set by the
well depth $V_0$ and the interaction strength $U$, both measured against the
kinetic energy. At $\phi=20^\circ$, each well has threefold ($C_3$) rather than sixfold symmetry, which the molecules inherit.

Theoretically, this problem admits a strong coupling description in the limit of large $a_M$, when the inter-moir\'e site Coulomb repulsion is significantly suppressed relative to the onsite repulsion~\cite{Artificial_Atoms}.
In this limit, one may first solve the problem of three interacting electrons in a single confining moir\'e potential well, where one finds a Wigner molecule with tri-lobed charge density as the on-site ground state~\cite{Artificial_Atoms,YannouleasLandman2023,Khalifa2026SciPost}.
The spin-state of each Wigner molecule is set by $\lambda$, the ratio of the local Coulomb energy $E_C$ to the (harmonic) confinement potential level spacing $\hbar \omega$.
For $\lambda < \lambda^* \approx 4.35$, the molecule has low-spin $S = 1/2$ and an orbital angular momentum $\ell = \pm 1$ degree of freedom, while for $\lambda > \lambda^*$ the molecule has high spin $S = 3/2$ and $\ell = 0$~\cite{Artificial_Atoms}.
An effective Hamiltonian describing the lattice of Wigner molecules then may be obtained by perturbative consideration of the electron hopping across molecules and intra-molecular interactions~\cite{Khalifa2026SciPost,veitas2026fluctuationdrivenchiralferromagnetism}.

Experimentally, however, the situation is less clear, as the experimental realizations~\cite{Li2024Science,Liu2026NSR} may not be in the large $a_M$ limit.
In particular, inter-site interactions may be strong enough to re-organize the effective low-energy degrees of freedom, or invalidate such a description altogether.
To directly address this regime, we choose a reference set of parameters (Table~\ref{tab:params}) that correspond to the twisted-WS$_2$ experiment~\cite{Li2024Science}.
We note that $\lambda=5.17$ for these parameters, implying a strong coupling prediction of $S = 3/2$ molecules, which we will find to be strongly violated.
To test the robustness of our conclusions, we repeat the analysis at a second point, where both the potential and the interaction are weaker relative to the kinetic energy.

\begin{table}[tb]
\centering
\begin{tabular}{lcccccc}
\toprule
 & $V_0$ (meV) & $m^\ast/m_e$ & $E_0$ (meV) & $V_0 / E_0$ & $U$ & $\lambda$\\
\midrule
reference & $10.3$ & $0.9$ & $0.88$ & $11.7$ & $33.3$ & $5.17$\\
second & $10.5$ & $0.7$ & $1.13$ & $9.3$ & $25.9$ & $4.26$\\
\bottomrule
\end{tabular}
\caption{The two parameter points. Both use $a_M=9.8$~nm, $\phi=20^\circ$, and
$\epsilon_r=5$. $\lambda=U/\sqrt{\omega}$ is the Wigner parameter of an isolated
harmonic well~\cite{Artificial_Atoms}, with
$\omega=4\pi(V_0\cos\phi)^{1/2}$.}
\label{tab:params}
\end{table}

The variational wavefunction is a self-attention neural network of the Psiformer
type~\cite{vonGlehn2023}, adapted to periodic solids~\cite{qxc3-bkc7,periodicwave_github};
its architecture and the hyperparameters of its training are listed in
Appendix~\ref{app:defs} (Table~\ref{tab:hyper}). We train the network by minimizing the variational energy. Each optimization
starts from a random initialization in a fixed sector of the total spin projection, $\Sztot=\tfrac12\sum_i\sigma_i$.
We consider the lowest sector, $\Sztot=0$; one-third magnetization, $\Sztot=N_e/6$; and the fully polarized state, $\Sztot=N_e/2$.
For the $N_e=36$ electrons on $N_{\rm site}=12$ sites studied here, these are $\Sztot=0$, $6$, and $18$.

\section{Beyond the Wigner-molecule crystal}
\label{sec:beyond}

Having introduced the model, we first ask whether the Wigner-molecule crystal is the ground
state. To follow the charge of each molecule, we assign every electron of a sampled
configuration to the Wigner--Seitz cell of the nearest potential minimum and count the
electrons in each cell. This gives the site charges $Q_A$ [Fig.~\ref{fig:cg}(a)], which
Sec.~\ref{sec:method} develops into a full readout of the wavefunction. In a Wigner-molecule
crystal, $Q_A=3$ at every site.

\subsection{Allowing spin fluctuations releases the charge}
\label{sec:charge}

The fully spin-polarized state comes closest to the crystal.
A configuration with every
molecule at three electrons is the norm, and the variance of the site charge,
$\var(Q)=\av{(Q_A-3)^2}=\sum_QP(Q)\,(Q-3)^2$, where $P(Q)$ is the probability that a site holds $Q$ electrons and $\av{Q}=3$ at every site, is
0.19 [Fig.~\ref{fig:cg}(d)]. It is not, however, the ground state. The $\Sztot=0$ state lies
23.6~meV lower (0.65~meV per electron), whereas
the entire low-spin manifold from $\Sztot=0$ to $\Sztot=6$ spans only 0.88~meV
(Table~\ref{tab:energy}). The energy in the total spin sectors is ordered as $E(\Sztot=0)<E(\Sztot=N_e/6)\ll E(\Sztot=N_e/2)$. This confirms the energy ordering found in Ref.~\cite{Li2025CommPhys} and
extends it to the lowest spin sector which has not been studied before.

Crucially, lifting the constraint of full spin-polarization changes the charge, not only the spin.
In the ground state, a site holds two electrons with probability 0.20 and four with 0.18, and a configuration with all twelve molecules at three electrons rarely occurs [Fig.~\ref{fig:cg}(d)]. The
variance of the site charge, 0.43, is more than twice that of the polarized state and retains
half of its value in the non-interacting system with the same potential and the same cells. The moir\'e potential pins the density to the same pattern at
every site, a lattice of identical triangles all oriented the same way, both in the
ground state, $\Sztot=0$ [Fig.~\ref{fig:cg}(b)], and in the fully polarized state, $\Sztot=N_e/2$
[Fig.~\ref{fig:cg}(c)]. The two differ only within each molecule, where in the ground state the three electrons are no longer resolved as three density maxima localized on the corner of a triangle. The triangle fills in, and the density at the minimum rises.
The $\Sztot=N_e/6$ state, in which the spins are polarized as far as possible without breaking the singlet pair of any molecule, has the same fluid charge as the ground state. What distinguishes the
fluid states from the polarized one is therefore not the arrangement of the spins, but the
singlet pairing of electrons within a molecule. In this sense, the Wigner-molecule crystal does
not survive once the spins are free: the ground state keeps the lattice of identical
triangles, but not its fixed charge.

\subsection{A metastable crystal on the way to the ground state}
\label{sec:path}

\begin{figure*}[t]
\includegraphics{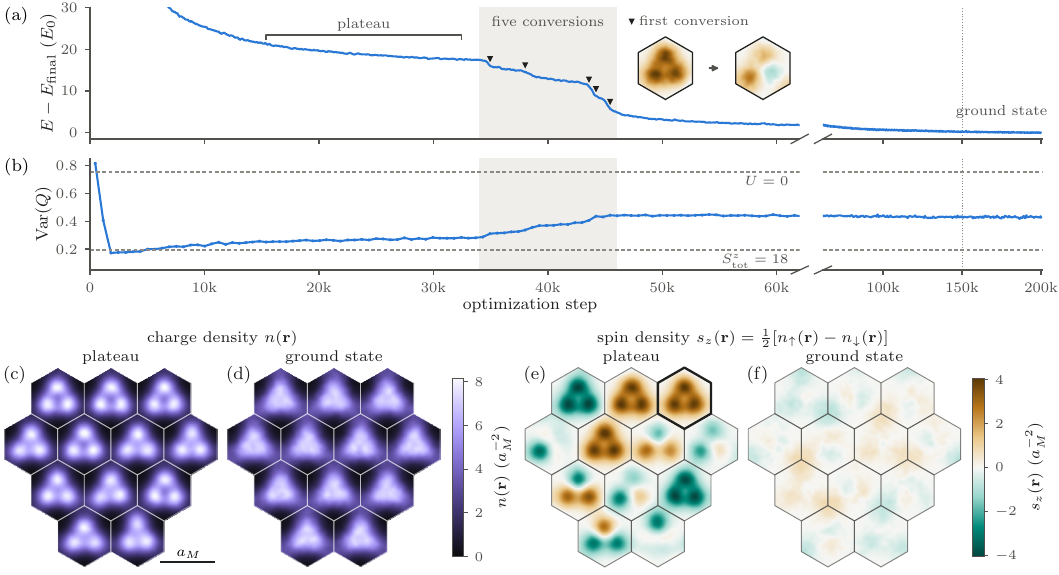}
\caption{Optimization of the $\Sztot=0$ state at the reference point, read through the
coarse-grained wavefunction at 332 checkpoints. (a)~The variational energy relative to its
final value (running average over 201 steps); triangles mark the conversions of the five
quartets, and the bracket marks the plateau checkpoints averaged in (c) and (e). Inset: the
spin density in the cell of the molecule that converts first, outlined in (e), over the four
checkpoints before (steps $3.18$--$3.37\times10^4$) and after ($3.49$--$3.68\times10^4$) its
conversion, on the color scale of (e),(f): a quartet, $\av{S^z_A}=1.36$, becomes a doublet,
$\av{S^z_A}=0.45$. (b)~The
site-charge variance, together with the non-interacting value and that of the
fully polarized state (dashed). Over the same conversions, the NN spin correlation turns from
ferromagnetic on the plateau to antiferromagnetic in the ground state.
Shading indicates the conversions, and the dotted line marks the step at which spin-exchange
moves were added to the sampling. The horizontal axis is broken at $6.2\times10^4$ steps.
(c),(d)~The charge density $n(\rr)$ and (e),(f)~the spin density
$s_z(\rr)=\tfrac12[n_\uparrow(\rr)-n_\downarrow(\rr)]$ on the twelve sites, drawn as in
Fig.~\ref{fig:cg}(a), on the plateau [(c),(e); 29 checkpoints between steps $1.5$ and
$3.3\times10^4$] and in the ground state [(d),(f); the last nine checkpoints]. On the plateau,
the three electrons of every molecule are resolved, as in the fully polarized state
[Fig.~\ref{fig:cg}(c)], and the spins form a static pattern: the five quartets carry
$|\av{S^z_A}|=1.40$--$1.46$, and the other seven molecules $0.41$--$0.47$. In the ground state,
the triangles fill in and the static moments nearly vanish. We use a gaussian smoothing of width
$0.035\,a_M$.  }
\label{fig:path}
\end{figure*}

The same conclusion emerges from the optimization itself. A trained network is the end point
of an optimization, and following the site charges and spins along the way reveals how the network gets there (Fig.~\ref{fig:path}). Within the first
percent of its $2\times10^5$ steps, the network has placed the electrons into the twelve
wells. It then settles onto a long plateau of slowly decreasing energy
[Fig.~\ref{fig:path}(a)], and the sampled configurations, coarse-grained and in real space,
show what the plateau state is [Fig.~\ref{fig:path}(b)--(f)].

The plateau state is a metastable crystalline state in the variational landscape, closely resembling the
Wigner-molecule crystal in character.
Its charge is nearly frozen, with a site-charge variance close to that of the fully polarized state [Fig.~\ref{fig:path}(b)], and, as in that state, the three electrons of every molecule are resolved in the density [Fig.~\ref{fig:path}(c)].
Its spins are static and pinned to the sites
[Fig.~\ref{fig:path}(e)], in the sense that every molecule carries a nonzero
expectation value of its spin projection, $\av{S^z_A}$, which we call its static moment: five of the twelve molecules are fully polarized quartets, with
$|\av{S^z_A}|\approx\tfrac32$, the other seven are doublets, with $|\av{S^z_A}|\approx\tfrac12$, and the neighboring moments tend to be parallel.
This is a broken-symmetry state of the mean-field kind, with the preference for parallel spins that exchange produces at the Hartree--Fock level.

The plateau ends in five discrete steps, one molecule at a time. At each step, one quartet
becomes a doublet, the singlet core plus one rim electron: within
about $10^3$ steps, the three parallel spins of the molecule give way to the static pattern of a
doublet, and its moment falls from $1.36$ to $0.45$ [inset of Fig.~\ref{fig:path}(a)]. In the
same step, the charge of that molecule begins to fluctuate, the charge variance of the lattice
grows, and the energy drops. Charge and spin thus change together, molecule by molecule. Each
conversion is a discrete, non-perturbative reorganization of the charge and spin density of one
molecule, reminiscent of a tunneling event between minima of the variational landscape.
By the end of the conversions, the static moments of the seven doublets have decayed as
well [compare Fig.~\ref{fig:path}(e) and (f)]: the spin of a molecule is no longer pinned to one
direction but fluctuates, so that $\av{S^z_A}$ averages to a small value, and the NN spin
correlations turn antiferromagnetic.
In the ground state, almost no static moment remains [Fig.~\ref{fig:path}(f)], and the triangles have
filled in [Fig.~\ref{fig:path}(d)].

The metastable crystal realizes both assumptions behind a description by fixed local moments:
every molecule holds a nearly fixed number of electrons, and each carries a static moment. The
variational landscape thus has a long-lived region where both hold, and the network leaves it
only by giving up both.

What, then, replaces the crystal, and which lattice model describes it? The site-charge
distribution alone cannot answer these questions. In the next section, we develop the readout
that does.

\section{Reading out the local structure: coarse-graining onto \moire{} sites}
\label{sec:method}

An effective lattice model needs more than the probability that a site holds a given number
of electrons. It needs the internal structure of a molecule of each charge, the spin state it
forms, and how neighboring molecules are correlated. To this end, we develop a readout that puts these
questions to the wavefunction directly by coarse-graining it onto the \moire{} sites.

We divide the plane into the Wigner--Seitz cells $\Omega_A$ of the potential minima
$\mathbf R_A$, and assign each electron to the cell that contains it. Every
configuration $X=(\rr_1\sigma_1,\ldots,\rr_{N_e}\sigma_{N_e})$ then defines a charge and a
spin for each site,
\begin{equation}
Q_A=\sum_i\chi_{\Omega_A}(\rr_i),\qquad
\SSS_A=\sum_i\chi_{\Omega_A}(\rr_i)\,\mathbf s_i ,
\label{eq:sites}
\end{equation}
where $\chi_{\Omega_A}$ is the indicator function of the cell, with $\sum_AQ_A=N_e$ and
$\sum_AS^z_A=\Sztot$ exactly [Fig.~\ref{fig:cg}(a)]. This construction carries the
partition of molecular charge densities into atomic basins~\cite{Bader1990,VDD2004} over to
the many-body wavefunction of a solid.

The site charges offer three levels of description. The first, the mean site charge,
equals three at every site, as fixed by the filling, and therefore carries no information about
the state. The second is the fluctuations of the site charges about this mean, described by the
probability $P(Q)$ that a site holds $Q$ electrons and, more completely, by the joint
distribution of all site charges.
Most of the physics sits in conditional
statistics. Fixing the charge of a site and asking for the density gives the internal
structure of a molecule of that charge; asking for its spin gives the multiplet it forms; and
fixing the charges at both ends of a bond connecting two molecules resolves the magnetism by charge environment.
These are statements about identifiable local objects, precisely the language in which an effective model is written.
We define each in turn.

\emph{Charge fluctuations.} With $\delta Q_A=Q_A-3$, charge conservation gives
$\sum_A\delta Q_A=0$ in every configuration. The site-charge correlation function
$C_{AB}=\av{\delta Q_A\,\delta Q_B}$ therefore obeys the sum rule
\begin{equation}
\sum_B C_{AB}=0 ,
\label{eq:sumrule}
\end{equation}
even in the absence of any physical preference: the charge a molecule gains or loses must be
made up elsewhere. We use $C_{AB}$ and this sum rule to describe how the fluctuations are
arranged. The charge variations relative to $Q=3$ are doublons, $d_A=(Q_A-3)_+$, and holes,
$h_A=(3-Q_A)_+$, so that $\delta Q_A=d_A-h_A$ and
$\sum_Ad_A=\sum_Ah_A\equiv n_{\rm exc}$ is the number of displaced electrons in a configuration. Here $(x)_+$ denotes the positive part, so that $d_A$ ($h_A$) measures the excess (deficit) of the charge $Q_A$ relative to 3 and vanishes otherwise.

\emph{Conditional densities.} For a site $A$ and charge $q$, the conditional density is
\begin{equation}
\begin{gathered}
n_A^{(q)}(\rr)=\av{\sum_i\delta(\rr-\rr_i)\,\Big|\,Q_A=q},\\
\int_{\Omega_A}n_A^{(q)}(\rr)\,d^2r=q ,
\end{gathered}
\label{eq:cond}
\end{equation}
the density of electrons around site $A$ in those configurations in which it holds $q$
electrons. It is the internal structure of a molecule of charge $q$. Where it extends beyond
$\Omega_A$, it shows how the neighbors respond. The unconditioned density in the cell is the
mixture $n_A(\rr)=\sum_qP_A(q)\,n_A^{(q)}(\rr)$. This identity is precisely why the density alone
cannot determine the site-charge distribution: very different distributions of $Q_A$ can
produce the same $n_A$. The differences $n_A^{(q)}-n_A^{(3)}$ show how a doublon ($q=4$) or a
hole ($q=2$) rearranges the charge relative to the three-electron molecule, and imposing a
second condition, on the charge $Q_B$ of a neighbor $B$, resolves how a molecule responds to its
environment. The orientation of a molecule is measured by the third angular harmonic of its
charge about the minimum,
$\psi_3^{(q)}=\av{Q_A^{-1}\sum_{i\in A}e^{3i\theta_i}\,|\,Q_A=q}$, where $\theta_i$ is the polar
angle of electron $i$ about $\mathbf R_A$.

\emph{Spin.} The spin sector requires additional care.
Longitudinal ($\av{S_A^zS_B^z}$) spin correlations follow
directly from the sampled spin labels, whereas the transverse ($\av{S_A^+S_B^-}$) parts follow from amplitude
ratios $R_{ij}(X)=\Psi(X^{ij})/\Psi(X)$, where $X^{ij}$ exchanges the spins of two electrons
$i$ and $j$ of opposite spin at fixed positions. The spin-transposition identity
$P_{ij}=\tfrac12+2\,\mathbf s_i\cdot\mathbf s_j$ relates these ratios to the transverse channel.
Let $T_{AB}(X)$ collect the ratios of all opposite-spin pairs with one electron in cell $A$ and
the other in cell $B$, added to both entries $AB$ and $BA$, and let
$m_A=\tfrac12\sum_{i\in A}\sigma_i$. Then
\begin{equation}
\av{\SSS_A\cdot\SSS_B}=\av{m_Am_B+\tfrac12\operatorname{Re}T_{AB}(X)+\tfrac{Q_A}{2}\,\delta_{AB}} ,
\label{eq:KAB}
\end{equation}
including the local $\av{\SSS_A^2}$ on the diagonal, without assuming spin-rotation invariance,
which the trained networks satisfy only approximately (Appendix~\ref{app:spin}). Restricting
the average to configurations with $Q_A=q$, or with $Q_A=q$ and $Q_B=q'$ on a bond, gives the
spin of a molecule of each charge and the magnetism of each charge environment. Since two
electrons can form only a singlet or a triplet, and three only a doublet or a quartet,
$\av{\SSS_A^2}$ at fixed $Q_A=2$ or 3 fixes the local multiplet content outright.

\section{Local structure of the ground state}
\label{sec:results}

\begin{figure*}[t]
\includegraphics{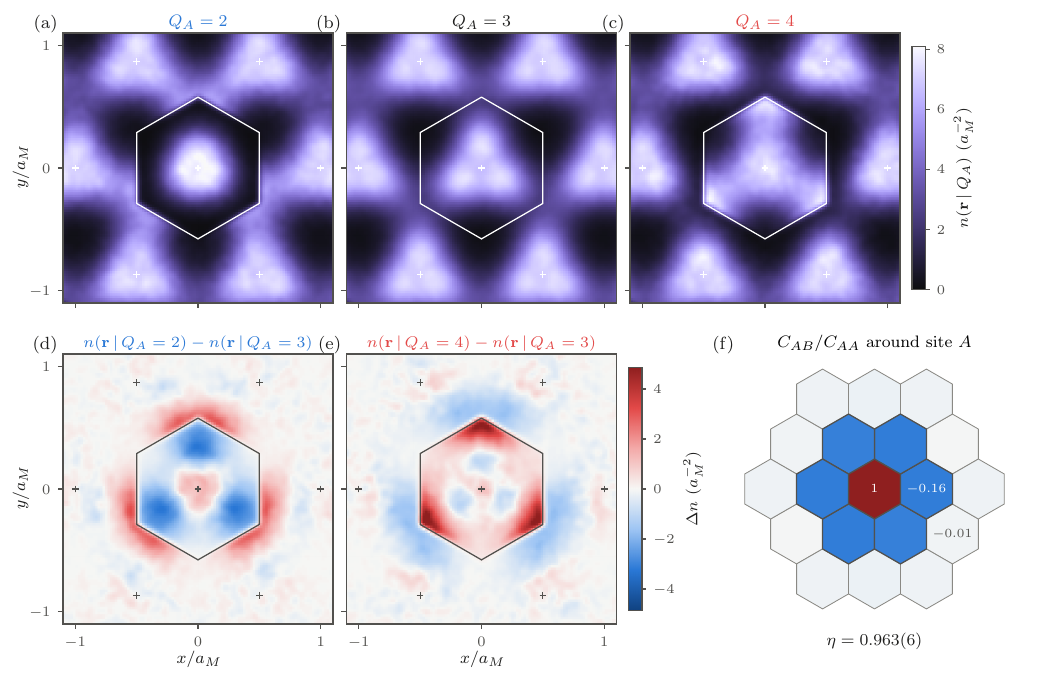}
\caption{The three kinds of Wigner molecule and the arrangement of the doublons and holes.
(a)--(c)~Density conditioned on the charge $Q_A$ of the central site ($\Sztot=0$, reference
point), averaged over the twelve sites, which are equivalent by translation; the maps computed
at each site separately are statistically identical. The hexagon marks the Wigner--Seitz cell
of the central site, and crosses mark the potential minima. (d)~The hole,
$n(\rr\,|\,Q_A=2)-n(\rr\,|\,Q_A=3)$: charge leaves the corners of the triangle (blue) and
appears just outside the cell (red), on the neighbors. (e)~The doublon,
$n(\rr\,|\,Q_A=4)-n(\rr\,|\,Q_A=3)$: the added charge sits at the corners of the triangle,
while the neighbors are depleted. Gaussian smoothing of width $0.025\,a_M$. (f)~The
site-charge correlation $C_{AB}/C_{AA}$ around a site $A$, averaged over the twelve sites. Each
of the six neighbors carries $-0.16$, so that a fraction $\eta=0.963(6)$ of the charge a
molecule gains or loses is made up on them [Eq.~(\ref{eq:eta})]. The outer cells are periodic
images of the remaining five sites.}
\label{fig:molecules}
\end{figure*}

We now apply the readout to the converged $\Sztot=0$ state at the reference point, then to the sector $\Sztot=N_e/6$, in which the emergent
moments are polarized, and finally propose the lattice model that these
observations motivate. The values behind each statement are collected in
Appendix~\ref{app:tables}.

\subsection{Molecules of two, three, and four electrons}
\label{sec:molecules}

The density hides the charge fluctuations because it averages over molecules of different
charge. Conditioning on the site charge separates them [Fig.~\ref{fig:molecules}(a)--(c)]. A
two-electron molecule is compact and centered on the minimum. A three-electron molecule forms
a triangle whose corners point toward the corners of its cell, as dictated by the $C_3$ well.
A fourth electron pushes weight out to the three corners, toward the edge of the cell. The
rms radii of the three are $0.24$, $0.31$, and $0.35\,a_M$ respectively. Weighted with their
probabilities, the three profiles recover the unconditioned one, so
that a single density is compatible with very different distributions of site charge. Every
molecule keeps the orientation imposed by its well, at all twelve sites, and the orientation
is pinned more firmly the more electrons the molecule holds. The molecules are therefore rigid
in shape and orientation, but fluid in charge.

\subsection{Bound doublon--hole pairs}
\label{sec:pairs}

Next, we ask how the charge fluctuations are arranged. By the sum rule
[Eq.~(\ref{eq:sumrule})], the
fraction made up by its six nearest neighbors,
\begin{equation}
\eta=-\frac{1}{C_{AA}}\sum_{B\in\NN(A)}C_{AB},
\label{eq:eta}
\end{equation}
equals $1$ if every fluctuation moves one electron between neighboring molecules, and $6/11$ if
the compensating charge is equally likely to sit on any other site. We find $\eta=0.963(6)$,
compared to $0.78$ without interactions [Fig.~\ref{fig:molecules}(f)]. The charge
fluctuations, $\var(Q)=0.43$ or 2.4 displaced electrons per configuration, therefore take the
form of tightly bound NN doublon--hole pairs.

The conditional densities reveal the same process in real space
[Fig.~\ref{fig:molecules}(d),(e)]. The charge a doublon gains sits at the three corners of its
cell, while the density just outside is depleted, marking the neighbors the electron came
from. A hole is the mirror image. Moreover, the molecules respond to the charge of their
neighbors, leaning toward a neighboring hole and away from a neighboring doublon. Relative to
a three-electron molecule, a doublon carries an extra charge $-e$ and a hole $+e$, and these
leans have precisely the signs of their mutual attraction and repulsion. The doublon responds
most: next to a hole, its loosely held extra charge piles up at the corner of their shared
edge, while the hole stays compact. A bound pair is therefore two distinct molecules held
together by their attraction, rather than an electron shared between two sites.

\subsection{The charge of a molecule sets its spin}
\label{sec:moments}

\begin{figure}[t]
\includegraphics{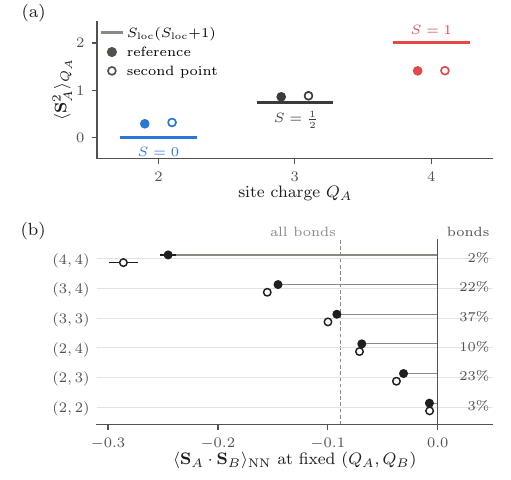}
\caption{Local moments and NN spin correlations set by the site charge ($\Sztot=0$). (a)~The local
spin at fixed site charge (points), compared to $S_{\rm loc}(S_{\rm loc}+1)$ from
Eq.~(\ref{eq:moment_rule}) (horizontal lines). (b)~The NN spin correlation resolved by the
charges at the two ends of the bond. The column on the right gives the fraction of NN bonds in
each class, and the dashed line indicates the average over all bonds. Filled (open) symbols
denote the reference (second) point.}
\label{fig:spin}
\end{figure}

We now turn to the spin sector. Conditioning the local spin on the site charge
[Fig.~\ref{fig:spin}(a)] shows that the three kinds of molecule carry different moments. A
two-electron molecule is a spin singlet with weight 0.85, a three-electron molecule a doublet
with weight 0.96, and a four-electron molecule a triplet with weight 0.69, the remainder being
mostly singlet. A single rule describes all three: two electrons form a closed
singlet core, while the remaining $Q-2$ occupy the rim with parallel spins,
\begin{equation}
S_{\rm loc}(Q)=\frac{Q-2}{2},\qquad Q=2,3,4 .
\label{eq:moment_rule}
\end{equation}
We emphasize that this rule names the dominant multiplet in each charge sector; it is not an
exact projection of the wavefunction. The core is robust: sites with five electrons, although much rarer than charge 3e or charge 4e molecules because of enhanced Coulomb repulsion, are nearly four times more common than sites that have lost a core electron.

Consequently, every change of the charge of a molecule also changes its local moment.
A hole removes a spin-$\tfrac12$ moment
from the lattice, whereas a doublon replaces it with a spin-one object. A spin-one doublon
requires its two rim electrons to occupy different orbitals, so that the rim is a shell of
several orbitals rather than a single one. The rule holds tightly for holes and
three-electron molecules, while the doublon is the soft exception: its triplet and singlet are
nearly degenerate, and the sizes of the molecules show why. Aligning the spins of a two- or
three-electron molecule forces its closed core to open, and the molecule increases in size. By contrast, a four-electron molecule keeps its size
whichever way its rim spins point, because its two rim electrons already sit in different
corners of the triangle. Flipping the rim spin of a doublon therefore costs almost no
rearrangement: the moment of the doublon is soft.

\subsection{Antiferromagnetism concentrated on the doublons}
\label{sec:afm}

The NN spin correlation is antiferromagnetic, $\av{\SSS_A\cdot\SSS_B}_\NN=-0.089$, and it
remains antiferromagnetic in every charge class of the bond [Fig.~\ref{fig:spin}(b)]. Its
magnitude, however, spans a factor of 30 across the classes. The antiferromagnetic
correlation is smallest on bonds touching a hole, as expected for a site that usually carries
no moment, intermediate on bonds between two three-electron molecules, and largest on bonds
touching a doublon: between two doublons it is 2.7 times larger than between two
three-electron molecules.
Indeed, bonds with a doublon make up a third of the lattice but carry half of the antiferromagnetic correlation.
Doublons and holes are thus not passive vacancies in a spin lattice: holes remove moments,
while doublons enhance the antiferromagnetic correlations of the moments that remain.

The spin correlations are also imprinted on the positions of the electrons. On bonds with
three electrons and $S^z_A=\pm\tfrac12$ at each end, two molecules approach each other more
closely when their moments are antiparallel, so that the wavefunction does not factorize into
a charge part and a spin part, even at fixed site charges. By contrast, the orientation of the
molecules plays no resolvable role: at fixed bond charges, the spin correlations do not depend
on how the neighboring triangles are oriented.

\subsection{Polarizing the emergent moments}
\label{sec:polar}

Equation~(\ref{eq:moment_rule}) makes a sharp counting prediction. If every molecule keeps its
singlet core, the total rim spin is
\begin{equation}
\sum_A S_{\rm loc}(Q_A)=\sum_A\frac{Q_A-2}{2}=\frac{N_e-2N_{\rm site}}{2}=\frac{N_e}{6}
\label{eq:saturation}
\end{equation}
in every charge configuration: the spin a doublon gains is exactly the spin its partner hole
loses. The emergent moments can therefore be fully polarized at $1/3$ magnetization without breaking a
single core, whereas the fully polarized state breaks all of them, and the charge fluctuations freeze.
For our $N_e = 36$ system, comparing $\Sztot=0$ with $\Sztot=6$ thus isolates the magnetism of the
emergent moments.

\begin{figure}[t]
\includegraphics{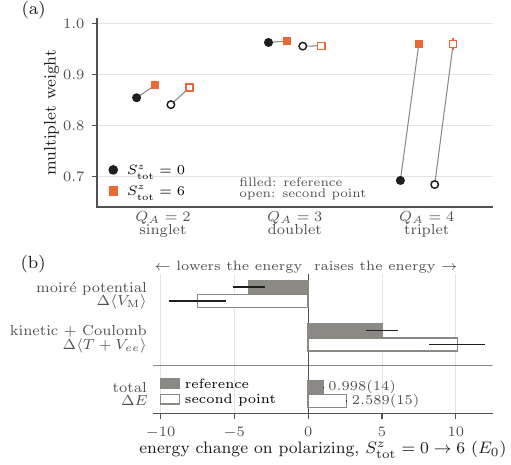}
\caption{Polarizing the emergent moments, $\Sztot=0$ to $\Sztot = 6$. (a)~Weight of the dominant multiplet of
each kind of molecule at $\Sztot=0$ (circles) and $\Sztot=6$ (squares). On polarizing, the
singlet weight of the hole and the doublet weight of the three-electron molecule change by at
most 0.04, whereas the triplet weight of the doublon rises from 0.69 to 0.96. For the doublon at
$\Sztot = 6$, $\av{\SSS_A^2}$ alone does not fix the triplet weight, because four electrons can
also form a quintet ($S=2$). The quintet weight is small but not measured in this sector, and
varying it between zero and its $\Sztot=0$ value gives a range of triplet weights
(Appendix~\ref{app:spin}); the square marks the middle of this range, which is narrower than the
symbol.
(b)~The energy change on
polarizing, $\Delta E$ (bottom), is the sum of the changes of the \moire{} potential energy and
of the kinetic plus Coulomb energy (top), in units of $E_0$ of each point. The \moire{}
potential energy is measured directly, while the kinetic plus Coulomb change follows by
subtraction. Filled (open) symbols and bars denote the reference (second) point; the
lines on the bars are statistical errors. The small cost of polarizing,
$\Delta E\approx1\,E_0$ at the reference point, is the residual of two larger, opposing
changes: the molecules contract, which lowers the \moire{} potential energy, at a slightly
larger cost in kinetic and interaction energy. $\Delta E$ is therefore not a single exchange
scale.
}
\label{fig:polar}
\end{figure}

We find that the $\Sztot=N_e/6$ state is the fully polarized end of the same molecular manifold. Its
site-charge distribution and the NN arrangement of its doublons and holes are those of the $\Sztot=0$ state
within statistical error, and its molecules are more compact. In the spin sector, the
response is concentrated on a single object [Fig.~\ref{fig:polar}(a)]: the multiplets of the
hole and of the three-electron molecule change little, while the doublon loses almost all of
its singlet admixture. Polarization thus changes the size of the
local moment, not only its direction, and the doublon is where this happens. In other words,
the local spin magnitude is a dynamical variable, not a fixed label attached to a site. The
transverse spin fluctuations that remain in the polarized state are again antiferromagnetic on
every bond class, and two thirds of them sit on bonds with a doublon.

Polarizing the emergent moments costs $0.998(14)\,E_0=0.88$~meV for the whole 36-electron
system. Crucially, this is not a single energy scale. As the molecules contract, the \moire{}
potential energy falls by $4.0(11)\,E_0$, while the kinetic and interaction energies rise by
slightly more [Fig.~\ref{fig:polar}(b)]; the splitting is the small residual of the two. Rather
than read it as an exchange constant, we use the local statistics to constrain a lattice model.

\subsection{Toward an effective lattice model}
\label{sec:lattice}

The local structure found above motivates an effective lattice model.
First, the two-electron core is closed and robust
(Sec.~\ref{sec:moments}), and it survives the polarization of the
emergent moments (Sec.~\ref{sec:polar}). We therefore treat it as inert
to leading order, so that site $A$ carries $n_A=Q_A-2$ rim electrons,
one on average. Second, within this description, a spin-one doublon
requires its two rim electrons to occupy different orbitals
(Sec.~\ref{sec:moments}). We thus retain two rim orbitals, $a=1,2$,
taken to be degenerate, and include a Hund's exchange that favors
the triplet. Third, the charge correlations indicate predominantly
local doublon--hole compensation (Sec.~\ref{sec:pairs}). This
motivates NN hopping and a repulsive NN density interaction, which
favors adjacent doublon--hole configurations. The unscreened Coulomb
interaction (Sec.~\ref{sec:intro}) motivates retaining this intersite
repulsion; its restriction to NN is a further approximation.
Combining these ingredients gives a two-orbital extended Hubbard
model of the rim,
\begin{equation}
\begin{aligned}
H_{\rm eff}={}&-\sum_{\langle AB\rangle}\sum_{ab\sigma}
\big(t^{ab}_{AB}c^\dagger_{Aa\sigma}c_{Bb\sigma}
+\mathrm{h.c.}\big)\\
&+\sum_A H_A
+V\sum_{\langle AB\rangle}(n_A-1)(n_B-1),\\
H_A={}&U\sum_a n_{Aa\uparrow}n_{Aa\downarrow}
+\big(U'-\tfrac12J_H\big)n_{A1}n_{A2}\\
&-2J_H\,\mathbf s_{A1}\cdot\mathbf s_{A2}
+J_P\big(P^\dagger_{A1}P_{A2}+\mathrm{h.c.}\big).
\end{aligned}
\label{eq:model}
\end{equation}
Here $c_{Aa\sigma}$ annihilates a rim electron with spin $\sigma$
in orbital $a$ of site $A$,
$n_{Aa\sigma}=c^\dagger_{Aa\sigma}c_{Aa\sigma}$,
$n_{Aa}=\sum_\sigma n_{Aa\sigma}$, $n_A=n_{A1}+n_{A2}$,
$\mathbf s_{Aa}$ is the spin of orbital $a$, and
$P_{Aa}=c_{Aa\downarrow}c_{Aa\uparrow}$.
The local interaction is the result of projecting the Coulomb interactions onto the two orbital subspace. It includes intra-orbital repulsion $U$,
inter-orbital repulsion $U'$, Hund's exchange $J_H>0$, and a real
pair-hopping amplitude $J_P$.
We leave $U$, $U'$, $J_H$, and $J_P$
independent. In particular, we do not impose the Kanamori relations
$U'=U-2J_H$ and $J_P=J_H$
\cite{Kanamori1963,Georges2013,Ralko_pinball}. Orbital degeneracy alone does not
establish these relations, and the symmetry and interaction matrix
elements of the active rim orbitals remain to be established.
Equation~(\ref{eq:model}) is therefore a minimal interaction ansatz,
rather than the most general local Hamiltonian. With one electron in four spin-orbitals per
site, the rim is quarter filled.

The dominant local multiplets can reproduce the moment rule of
Eq.~(\ref{eq:moment_rule}). A hole ($n_A=0$) is the bare spin-singlet
core, and a three-electron molecule ($n_A=1$) carries spin
$\tfrac12$.
For a doublon ($n_A=2$), the inter-orbital spin-triplet has
energy $U'-J_H$, while the inter-orbital spin-singlet has energy
$U'+J_H$ and the two same-orbital singlet combinations have
energies $U\pm J_P$. The spin-triplet is therefore the lowest doublon
multiplet when $U'-J_H<U-|J_P|$, which we take as a constraint
motivated by the observed triplet preference. In the lattice, hopping-induced charge fluctuations
and intersite entanglement allow both singlet and triplet doublon
weights, in competition with this local preference. Once the rim
is fully polarized, as at $\Sztot=6$, the two rim electrons of a
doublon must be parallel, capturing the leading response in
Fig.~\ref{fig:polar}(a).

For $V>0$, the NN interaction lowers the energy of an adjacent
doublon--hole pair and raises that of two adjacent doublons or two adjacent holes.
These signs are consistent with the attraction and repulsion
inferred from the leans of the molecules
(Sec.~\ref{sec:pairs}).
In this minimal model intersite spin correlations are generated
by the same hopping that produces charge fluctuations.
Because Hund's exchange favors the triplet on a doublon, the antiferromagnetic correlations found here constrain the hopping structure and interaction parameters.
The molecular orientation, pinned by the $C_3$ well and without a resolvable effect on the spin correlations (Sec.~\ref{sec:afm}), enters as a fixed
background that constrains the hopping matrices $t^{ab}_{AB}$.

The local statistics presented here provide natural benchmarks
for this model. The site-charge distribution, the compensation
fraction $\eta$, and the doublon multiplet weights constrain the
hopping structure and the dimensionless ratios $U/t$, $U'/t$,
$V/t$, $J_H/t$, and $J_P/t$, where $t$ denotes a characteristic
hopping scale. The charge-resolved NN spin correlations and the
response to polarization provide further tests. Because the
polarization energy is the small residual of larger structural
changes (Sec.~\ref{sec:polar}), we regard these local statistics
as the primary benchmarks, with energy differences providing
complementary constraints on the overall scale and on the effects
of molecular relaxation. Identifying the active rim orbitals,
deriving their interaction and hopping matrix elements from the
continuum description, and solving the resulting model on
lattices large enough to address its ground state are natural
next steps, which we leave to a future study.

\section{Robustness at a second parameter point}
\label{sec:robustness}

To separate properties of the state from those of a particular parameter choice, we repeated
the $\Sztot=0$ and $\Sztot=6$ calculations at the second point of Table~\ref{tab:params}. The
picture carries over as a whole (open symbols in Figs.~\ref{fig:spin} and \ref{fig:polar}). The
molecules are pinned with a common orientation and fluid in charge, and their charge
fluctuations are NN doublon--hole pairs to the same degree, $\eta=0.954(12)$ compared to
$0.963(6)$. The charge of a molecule sets its moment according to
Eq.~(\ref{eq:moment_rule}). The NN spin correlations are antiferromagnetic in every charge
class, and doublon bonds carry half of them at one third of the bonds. Polarizing the emergent
moments again acts on the doublon and contracts the molecules, at an energy cost that is again
the residual of larger, opposing changes. Notably, at this point $\lambda=4.26$ lies below the
single-molecule $\lambda^*$, where an isolated harmonic molecule is itself a doublet; the
singlet core and the fluid charge of the lattice state therefore do not hinge on the level
crossing of an isolated molecule.

The differences are quantitative and in the expected direction. The lighter mass gives the
kinetic energy more weight against both the potential and the interaction ($\lambda=4.26$
compared to $5.17$), and the state is more fluid: the charge variance rises to 0.49, the
molecules are larger and less firmly oriented, and polarizing the emergent moments costs
$2.6\,E_0$ (2.9~meV). The additional fluctuations are of the same local kind, since $\eta$ is
unchanged within its error; in other words, the second point has more doublon--hole pairs but
still tightly bound.

\section{Conclusion}
\label{sec:discussion}
Continuum neural wavefunctions have great variational power to approximate the ground states of systems with competing energy scales.
The huge number of variational parameters makes it challenging to form an understanding of the organizing physics behind these systems.
The readout developed in this work puts physical understanding onto that solution.
By coarse-graining the wavefunction onto site-resolved charge and spin statistics, it reveals the effective degrees of freedom of the state.

Applied to \moire{} semiconductors at $\nu=3$, where the potential strength, the electron-electron interactions, and the kinetic energy are all comparable, the NQS reveals a ground state vastly different from the mean-field ferromagnetic Wigner molecule crystal. It is spin unpolarized and its charge fluctuates strongly. The readout shows that this state has a simple local organization. Its molecules are rigid in orientation but fluid in charge, each built from an inert singlet core and a spin-carrying rim, and charge moves between neighbors as bound doublon--hole pairs that also carry spin. These emergent low-energy degrees of freedom point to a two-orbital extended Hubbard model of the rim, whose parameters are constrained by the local statistics.

Several directions follow from this work. Doping away from $\nu=3$ filling is particularly interesting, since added or removed carriers enter a state that already hosts doublons and holes with spin attached, and the effective model, which can be solved on much larger lattices than the continuum calculation allows, can serve as a companion in this regime. The method also extends nicely to other \moire{} TMD systems, in particular the generalized Wigner crystals at fractional fillings~\cite{Regan2020}. A recent study of their response to a magnetic field has indicated that these systems may have a nontrivial magnetic structure \cite{wolff2026magneticfieldresponsegeneralizedwigner}, which a wavefunction-level readout could address directly.

\begin{acknowledgments}
The authors acknowledge helpful discussions with Senthil Todadri, Tomohiro Soejima and Ahmed Abouelkomsan.
The computations in this work were done using Bridges-2 at PSC and SDSC Expanse cluster through allocation PHY250361 from the Advanced Cyberinfrastructure Coordination Ecosystem: Services \& Support (ACCESS) program~\cite{access}, which is supported by U.S. National Science Foundation grants No. 2138259, No. 2138286, No. 2138307, No. 2137603, and No. 2138296.
\end{acknowledgments}

\appendix

\section{Neural-network wavefunction, sampling, and errors}
\label{app:defs}

\subsection{Network and optimization}
For completeness, we summarize the neural-network wavefunction, which follows
Ref.~\cite{qxc3-bkc7}, and its optimization. The hyperparameters are listed in
Table~\ref{tab:hyper}.

Each electron carries a position and a spin label, $x_i=(\rr_i,\sigma_i)$ with
$\sigma_i=\pm1$. The wavefunction of the $N$ electrons in the periodic cell is a sum of $K$
generalized Slater determinants multiplied by a Jastrow factor,
\begin{equation}
\Psi(x_1,\dots,x_N)=e^{J}\sum_{k=1}^{K}
\det\big[\phi^k_j(x_i;\{x_{/i}\})\big]_{i,j=1}^{N}.
\label{app:eq:psi}
\end{equation}
In an ordinary Slater determinant, the orbital $\phi_j$ of electron $i$ depends only on
$x_i$. Here it also depends on the coordinates $\{x_{/i}\}$ of all other electrons,
symmetrically under their permutations, which generalizes the backflow transformation.
Exchanging two electrons still swaps two rows of every determinant, so $\Psi$ is
antisymmetric by construction. All $N$ electrons share one $N\times N$ determinant, with the
spin entering as an input to the orbitals; the ansatz therefore fixes $\Sztot$.

The Jastrow factor is a sum over pairs,
\begin{equation}
J=\sum_{i<j}u_{\sigma_i\sigma_j}(r_{ij}),\qquad
u_{\sigma\sigma'}(r)=-\frac{c_{\sigma\sigma'}\,\alpha_{\sigma\sigma'}^2}{\alpha_{\sigma\sigma'}+r},
\label{app:eq:jastrow}
\end{equation}
with $c_{\sigma\sigma'}=U/3$ for parallel and $c_{\sigma\sigma'}=U$ for antiparallel spins,
and two learned lengths $\alpha_{\sigma\sigma'}$. Because $u_{\sigma\sigma'}'(0)=c_{\sigma\sigma'}$,
the factor cancels the divergence of the Coulomb term $U/r_{ij}$ in the local energy when two
electrons meet, i.e., it imposes the two-dimensional cusp conditions of
Eq.~\eqref{eq:hamiltonian}. The two cases differ because the determinants vanish at the
coalescence of parallel spins and stay finite for antiparallel spins~\cite{qxc3-bkc7}. The
distance $r_{ij}$ is a smooth periodic version of $|\rr_i-\rr_j|$ that reduces to it at short
distances.

The orbitals are produced by a neural network that treats every electron identically. Each
electron enters through periodic functions of its position ($\sin$ and $\cos$ of
$\mathbf G_a\cdot\rr_i$, with $\mathbf G_a$ the reciprocal vectors of the cell) together with
its spin, which makes $\Psi$ periodic in the cell. These inputs are mapped to a vector for
each electron, which four layers then refine. In each layer, a self-attention step updates the
vector of every electron by a weighted sum over the vectors of all electrons, with weights that
depend on the configuration; this is how the orbital of one electron learns where the others
are. A small perceptron then acts on each electron's vector separately. Because the same
operations are applied to every electron and the attention sums over all of them, the result
is symmetric under permutations of the other electrons, as Eq.~\eqref{app:eq:psi} requires.
The complex orbitals are linear projections of the final vectors.

The variational energy is estimated as the mean of the local energy $H\Psi/\Psi$ over
configurations sampled from $|\Psi|^2$, with the Coulomb interaction evaluated by Ewald
summation with a neutralizing background. It is minimized with the Kronecker-factored
approximate-curvature (KFAC) optimizer~\cite{MartensGrosse2015,PhysRevResearch.2.033429},
starting from a random initialization. The same architecture is used for all states and at
both parameter points.

\begin{table}[tb]
\centering
\footnotesize
\setlength{\tabcolsep}{4pt}
\begin{tabular}{ll}
\toprule
\multicolumn{2}{l}{\textit{Network}}\\
Self-attention layers & 4\\
Heads $\times$ head width & $4\times32$\\
Perceptron width & 128, two layers per block\\
Determinants & 2\\
Jastrow factor & electron--electron, Coulomb cusps\\
Parameters & $1.66\times10^6$\\
\midrule
\multicolumn{2}{l}{\textit{Optimization}}\\
Optimizer & KFAC~\cite{MartensGrosse2015,PhysRevResearch.2.033429}\\
Learning rate & $\epsilon_0=3\times10^{-3}$\\
Schedule & linear warm-up over $10^4$ steps,\\
 & then $\epsilon_0/[1+(t-10^4)/10^4]$\\
Damping & $10^{-3}$\\
Norm constraint & $10^{-3}$\\
Covariance decay & $0.95$\\
Momentum & $0$\\
Local-energy clipping & $5\times$ mean absolute deviation\\
 & from the median\\
Walkers & 2048 \\
Steps & $2\times10^5$ ($10^5$ for $\Sztot=18$)\\
\midrule
\multicolumn{2}{l}{\textit{Sampling}}\\
Steps per iteration & 30\\
Proposal & Gaussian, width adapted\\
 & every 100 steps\\
Burn-in & 700 steps\\
\bottomrule
\end{tabular}
\caption{Architecture of the neural-network wavefunction and hyperparameters of its
optimization and sampling. $t$ is the optimization step.}
\label{tab:hyper}
\end{table}

\subsection{Sampling}

All states are optimized for $2\times10^5$ steps, except the $\Sztot=18$ control ($10^5$ steps),
which we use only qualitatively. At the reference point, each sector is sampled at nine
checkpoints of 2048 configurations. At the second point, the $\Sztot=0$ state is sampled at three
checkpoints of 2048 configurations and the $\Sztot=6$ state at six checkpoints of 1024. For
Sec.~\ref{sec:path}, the $\Sztot=0$ reference-point state is sampled at the 332 checkpoints saved
during its optimization, each with 2048 configurations. Up to step $1.5\times10^5$, the spin
label of each electron is fixed and only positions are sampled; afterward, spin-exchange moves
are added, which changes none of the coarse-grained observables.

\subsection{Statistical errors}

Walkers are assigned to 16 fixed blocks. The quoted error is the larger of the
block-to-block and the checkpoint-to-checkpoint standard errors; the latter captures
fluctuations common to all walkers of one checkpoint. With fewer than four checkpoints, only
the block error is available. None of these errors includes the variational bias of an
independently optimized state, which is why we interpret small differences between spin
sectors only when they recur at both parameter points. Along the optimization, only the block
error of each checkpoint is available; every change described in Sec.~\ref{sec:path} is many
times larger and extends over several checkpoints.

\section{Partition, conditional densities, and charge correlations}
\label{app:partition}

\begin{table*}[tb]
\centering
\begin{tabular}{lccccc}
\toprule
 & \multicolumn{3}{c}{reference point} & \multicolumn{2}{c}{second point}\\
Observable & $\Sztot=0$ & $\Sztot=6$ & $\Sztot=18$ & $\Sztot=0$ & $\Sztot=6$\\
\midrule
$P(2)$ & $0.19780(60)$ & $0.19567(56)$ & $0.09090(95)$ & $0.2172(14)$ & $0.2197(13)$\\
$P(3)$ & $0.6064(11)$ & $0.6106(12)$ & $0.8170(19)$ & $0.5650(24)$ & $0.5624(25)$\\
$P(4)$ & $0.18407(54)$ & $0.18246(59)$ & $0.08875(88)$ & $0.2003(11)$ & $0.2006(14)$\\
$\var(Q)$ & $0.4293(15)$ & $0.4237(19)$ & $0.1932(24)$ & $0.4881(30)$ & $0.4905(28)$\\
$\av{n_{\rm exc}}$ & $2.4328(71)$ & $2.4042(86)$ & $1.118(12)$ & $2.716(15)$ & $2.730(14)$\\
$P(Q_A=3\ \text{for all }A)$ & $0.01134(83)$ & $0.01291(86)$ & $0.2731(53)$ & $0.00439(91)$ & $0.0065(12)$\\
$\eta$ & $0.9626(63)$ & $0.9667(64)$ & $0.9861(89)$ & $0.954(12)$ & $0.9708(99)$\\
radius, $Q=2$ & $0.24266(30)$ & $0.24113(34)$ & $0.25755(75)$ & $0.25142(41)$ & $0.24825(64)$\\
radius, $Q=3$ & $0.30732(16)$ & $0.30714(13)$ & $0.29708(19)$ & $0.31321(22)$ & $0.31263(27)$\\
radius, $Q=4$ & $0.35333(19)$ & $0.35289(21)$ & $0.35107(43)$ & $0.35440(36)$ & $0.35346(38)$\\
$|\psi_3|$, $Q=2$ & $0.0646(42)$ & $0.0565(35)$ & $0.1527(89)$ & $0.0590(58)$ & $0.0575(58)$\\
$|\psi_3|$, $Q=3$ & $0.2065(19)$ & $0.2043(16)$ & $0.2152(26)$ & $0.1795(21)$ & $0.1720(29)$\\
$|\psi_3|$, $Q=4$ & $0.2544(29)$ & $0.2507(22)$ & $0.2238(57)$ & $0.2238(36)$ & $0.2251(37)$\\
\bottomrule
\end{tabular}
\caption{Charge sector at the reference point ($\Sztot=0$, $6$, $18$) and the second point
($\Sztot=0$, $6$). At the reference point without interactions, $\var(Q)=0.75$ and $\eta=0.78$. Second-point
$\Sztot=0$ errors are walker-block errors only.}
\label{tab:charge}
\end{table*}

\subsection{Cell and partition}

With the \moire{} lattice vectors $\mathbf a_1=a_M(1,0)$ and
$\mathbf a_2=a_M(-\tfrac12,\tfrac{\sqrt3}{2})$, the periodic cell is spanned by
\begin{equation}
\begin{aligned}
\mathbf L_1&=2\mathbf a_1+4\mathbf a_2=2\sqrt3\,a_M\,(0,1),\\
\mathbf L_2&=2\mathbf a_1-2\mathbf a_2=2\sqrt3\,a_M\,\bigl(\tfrac{\sqrt3}{2},-\tfrac12\bigr).
\end{aligned}
\end{equation}
The two vectors have equal length $2\sqrt3\,a_M$ and enclose $120^\circ$, so that the cell
holds twelve sites and is $C_6$ symmetric. Distances in the cell follow the minimum-image
convention: a displacement $\rr$ is represented by the shortest of its periodic images,
\begin{equation}
|\rr|_{\rm mi}=\min_{n_1,n_2\in\mathbb Z}\bigl|\rr+n_1\mathbf L_1+n_2\mathbf L_2\bigr| .
\end{equation}
This shortest image is unique whenever $|\rr|_{\rm mi}$ is smaller than the minimum-image
radius $|\mathbf L_1|/2=\sqrt3\,a_M$, half the shortest supercell vector; at or beyond it,
two or more images can tie. Each electron is assigned to the potential minimum $\mathbf R_A$
that minimizes $|\rr_i-\mathbf R_A|_{\rm mi}$, which coincides with the Wigner--Seitz hexagon
of that minimum for every sampled electron. At $\phi=20^\circ$, the minima sit at the lattice
points. Site separations fall into shells at $a_M$, $\sqrt3a_M$, and $2a_M$ with six, three,
and two neighbors; only the first lies strictly inside the minimum-image radius.

\subsection{Doublons, holes, and charge correlations}

The doublon and hole counts are $d_A=(Q_A-3)_+$ and $h_A=(3-Q_A)_+$, so that $\delta Q_A=d_A-h_A$ and
$\sum_Ad_A=\sum_Ah_A\equiv n_{\rm exc}$ in every configuration. The NN correlation
$C_\NN=\av{\delta Q_A\,\delta Q_B}_\NN$ is the second moment of the joint distribution
$P_\NN(q,r)$ of the charges at the two ends of a NN bond,
$C_\NN=\sum_{q,r}P_\NN(q,r)\,(q-3)(r-3)$, and splits into a doublon--hole part and two
parts between like charges, $C_\NN=C^{dd}_\NN+C^{hh}_\NN-2C^{dh}_\NN$, where $C^{xy}_\NN$ is the NN
covariance of $x_A$ and $y_B$. At the reference point ($\Sztot=0$), these are $-0.0166$,
$-0.0140$, and $+0.0192$: the doublon--hole correlation gives 56\% of $C_\NN=-0.0689$, and the
anticorrelation of two doublons or two holes the rest. The sum rule leaves $1-\eta=0.04$ of $C_{AA}$ on the
other five sites.

\section{Spin estimators}
\label{app:spin}

\subsection{Full spin correlations from exchange ratios}

Here, we give the details of the estimator of Eq.~(\ref{eq:KAB}). The ratios
$R_{ij}$ are evaluated at the sampled positions with the two opposite spin labels exchanged.
Summing all entries of the local estimator
$K_{AB}(X)=m_Am_B+\tfrac12\operatorname{Re}T_{AB}(X)+\tfrac{Q_A}{2}\,\delta_{AB}$
gives the total-spin estimator
$\Sztot(\Sztot+1)+N_\downarrow+\operatorname{Re}\sum_{i\in\uparrow,j\in\downarrow}R_{ij}$. Each
configuration costs one network evaluation per opposite-spin pair, i.e., 324 at $\Sztot=0$. For a charge class, values are first averaged over
the sites or bonds of that class within each configuration and then over configurations.

\subsection{Local multiplets}

If $w_{S|q}$ denotes the weight of total spin $S$ on a site of charge $q$, then
$s_q\equiv\av{\SSS_A^2}_{Q_A=q}=\sum_Sw_{S|q}S(S+1)$ with $\sum_Sw_{S|q}=1$. Since two
electrons allow $S=0,1$ and three allow $S=\tfrac12,\tfrac32$, we obtain $w_{0|2}=1-s_2/2$ and
$w_{1/2|3}=(15/4-s_3)/3$ without any assumption, in any spin sector. Four electrons, however,
allow $S=0,1,2$, and $s_4$ alone does not fix three weights:
\begin{equation}
w_{1|4}=\frac{s_4}{2}-3w_{2|4},\qquad w_{0|4}=1-\frac{s_4}{2}+2w_{2|4}.
\label{app:eq:quintet}
\end{equation}
We take the small quintet weight $w_{2|4}$ from the distribution of $m_A$ at $\Sztot=0$ (0.004
at the reference point and 0.007 at the second point). At $\Sztot=6$, we let it range from zero
to that value, which gives the intervals quoted for the doublon.

\subsection{Spin-rotation symmetry of the ansatz}

At $\Sztot=0$, $\av{\SSS^2}=1.92$ rather than zero, and the NN ratio
$\av{\SSS_A\cdot\SSS_B}/\av{S^z_AS^z_B}$ is $2.7$ rather than the isotropic value of three,
uniformly across bond classes. This is why we measure the transverse channel directly.

\section{Tables of values}
\label{app:tables}

Here, we collect the values behind the statements of the main text. Sampling errors are given
in the last digits in parentheses. Bond classes are unordered pairs of site charges, and radii
are rms distances of a site's own electrons from its minimum, in units of $a_M$.

The energy decomposition of Table~\ref{tab:energy} proceeds as follows. With
$H=T+V_{\rm M}+V_{ee}$, the change of the variational energy between sectors splits exactly
into $\Delta E=\Delta\av{T}+\Delta\av{V_{\rm M}}+\Delta\av{V_{ee}}$. The \moire{} term is
evaluated directly on the sampled configurations, while the kinetic and interaction terms
together follow by subtraction. A harmonic estimate,
$\Delta\av{V_{\rm M}}\simeq N_e\av{V_{\rm M}-V_{\min}}\,\Delta\av{r^2}/\av{r^2}$, with
$\av{r^2}$ the mean-square distance of the electrons from their minima, reproduces the measured
change at both points.

\begin{table}[tb]
\centering
\footnotesize
\setlength{\tabcolsep}{2.4pt}
\begin{tabular}{lcccc}
\toprule
 & \multicolumn{2}{c}{reference point} & \multicolumn{2}{c}{second point}\\
 & $\Sztot=0$ & $\Sztot=6$ & $\Sztot=0$ & $\Sztot=6$\\
\midrule
$Q=2$, singlet & $0.8545(16)$ & $0.8791(18)$ & $0.8409(27)$ & $0.8745(36)$\\
$Q=3$, doublet & $0.9631(6)$ & $0.9659(5)$ & $0.9559(9)$ & $0.9564(13)$\\
$Q=4$, triplet & $0.692(3)$ & $0.954$--$0.967$ & $0.684(6)$ & $0.949$--$0.971$\\
\bottomrule
\end{tabular}
\caption{Weight of the dominant local multiplet. The $Q=2$ and $Q=3$ weights follow from
$\av{\SSS_A^2}$ without assumptions; the $Q=4$ weight uses the small-quintet estimate of
Appendix~\ref{app:spin}, and at $\Sztot=6$ it is the resulting interval.}
\label{tab:multiplets}
\end{table}

\begin{table}[tb]
\centering
\footnotesize
\begin{tabular}{lcccc}
\toprule
 & \multicolumn{2}{c}{reference point} & \multicolumn{2}{c}{second point}\\
Class & fraction & $\av{\SSS_A\cdot\SSS_B}$ & fraction & $\av{\SSS_A\cdot\SSS_B}$\\
\midrule
all & $1$ & $-0.08859(53)$ & $1$ & $-0.09693(93)$\\
$(4,4)$ & $0.022$ & $-0.2453(72)$ & $0.027$ & $-0.286(13)$\\
$(3,4)$ & $0.219$ & $-0.1453(12)$ & $0.224$ & $-0.1551(31)$\\
$(3,3)$ & $0.374$ & $-0.09180(98)$ & $0.324$ & $-0.0999(19)$\\
$(2,4)$ & $0.102$ & $-0.0691(15)$ & $0.117$ & $-0.0712(32)$\\
$(2,3)$ & $0.234$ & $-0.03113(75)$ & $0.241$ & $-0.0376(13)$\\
$(2,2)$ & $0.026$ & $-0.0076(17)$ & $0.033$ & $-0.0074(18)$\\
\midrule
doublon bonds & & $0.51\ (0.34)$ & & $0.55\ (0.37)$\\
\bottomrule
\end{tabular}
\caption{NN spin correlation $\av{\SSS_A\cdot\SSS_B}$ at $\Sztot=0$ by bond class, with the
fraction of NN bonds in each class. Rare charges outside $Q=2,3,4$ enter the all-bond row
only. The last row gives the share of the summed class correlation carried by bonds with at
least one doublon, with their fraction of bonds in parentheses.}
\label{tab:bonds}
\end{table}

\begin{table}[tb]
\centering
\footnotesize
\begin{tabular}{lcccc}
\toprule
 & \multicolumn{2}{c}{reference point} & \multicolumn{2}{c}{second point}\\
Class & $C^\perp$ & enh. & $C^\perp$ & enh.\\
\midrule
$(3,4)$ & $-0.01238(65)$ & $1.44$ & $-0.0149(12)$ & $1.49$\\
$(2,4)$ & $-0.0258(12)$ & $3.01$ & $-0.0205(22)$ & $2.05$\\
$(4,4)$ & $-0.0174(44)$ & $2.03$ & $-0.0263(50)$ & $2.63$\\
$(3,3)$ & $-0.00436(46)$ & $0.51$ & $-0.0063(10)$ & $0.63$\\
$(2,3)$ & $-0.00531(55)$ & $0.62$ & $-0.00594(80)$ & $0.59$\\
$(2,2)$ & $-0.00100(69)$ & $0.12$ & $-0.0018(14)$ & $0.18$\\
\bottomrule
\end{tabular}
\caption{Transverse NN correlation $C^\perp=\tfrac12\av{S_A^+S_B^-+S_A^-S_B^+}$ at $\Sztot=6$ by
bond class, and its enhancement: the class's share of the summed transverse correlation
divided by its fraction of bonds (fractions as in Table~\ref{tab:bonds}).}
\label{tab:transverse}
\end{table}

\begin{table}[tb]
\centering
\footnotesize
\begin{tabular}{lcc}
\toprule
 & reference point & second point\\
\midrule
$\Delta E$ & $0.998(14)$ & $2.589(15)$\\
$\Delta E$ (meV) & $0.880(12)$ & $2.935(17)$\\
$E(\Sztot=18)-E(\Sztot=0)$ & $26.73$ & ---\\
$\Delta\av{V_{\rm M}}$ & $-4.0(11)$ & $-7.5(19)$\\
$\Delta\av{T+V_{ee}}$ & $5.0(11)$ & $10.1(19)$\\
rms contraction (\%) & $0.144(48)$ & $0.286(82)$\\
$\Delta\av{n_{\rm exc}}$ & $-0.029(11)$ & $+0.015(21)$\\
$\Delta\var(Q)$ & $-0.0056(25)$ & $+0.0024(41)$\\
largest arrangement shift & $1.1$ & $1.3$\\
$\av{\SSS^2}$ at $\Sztot=0$ & $1.923(17)$ & $1.627(18)$\\
$\av{\SSS^2}$ at $\Sztot=6$ & $42.0787(55)$ & $42.0272(76)$\\
$\av{\SSS_A\cdot\SSS_B}/\av{S^z_AS^z_B}$, NN, $\Sztot=0$ & $2.715(19)$ & $2.68$\\
\bottomrule
\end{tabular}
\caption{Polarization and energetics, in units of $E_0$ of each point; differences are $\Sztot=6$ minus $\Sztot=0$ unless stated. The
arrangement shift is the largest change, in standard errors, among seven NN measures of
the arrangement of doublons and holes.}
\label{tab:energy}
\end{table}

\clearpage
\bibliographystyle{apsrev4-2}
\bibliography{references_arxiv}
\end{document}